\documentclass[letterpaper, 12pt]{article}

\usepackage{bejournal}          
\usepackage{mathptmx}
\usepackage{graphics}
\usepackage[authoryear,comma,longnamesfirst,sectionbib]{natbib}

\usepackage{dcolumn}
\usepackage{bm}
\usepackage{amsmath}
\usepackage{amsfonts}   
\usepackage{graphicx} 
\usepackage{url}
\usepackage{float}
\begin{document}

\title{Random Walk Picture of Basketball Scoring}
\author{Alan Gabel$^*$ and S. Redner\footnote{Center for Polymer Studies and Department of Physics, Boston University, Boston,
Massachusetts 02215, USA}} 
\originalmaketitle

\begin{abstract}
We present evidence, based on play-by-play data from all 6087 games from the 2006/07--2009/10 seasons of the National Basketball Association (NBA), that basketball scoring is well described by a continuous-time anti-persistent random walk. The time intervals between successive scoring events follows an exponential distribution, with essentially no memory between different scoring intervals. By including the heterogeneity of team strengths, we build a detailed computational random-walk model that accounts for a variety of statistical properties of scoring in basketball games, such as the distribution of the score difference between game opponents, the fraction of game time that one team is in the lead, the number of lead changes in each game, and the season win/loss records of each team.
\end{abstract}

\newpage

\section{Introduction}

Sports provide a rich laboratory in which to study competitive behavior in a
well-defined way.  The goals of sports competitions are simple, the rules are
well defined, and the results are easily quantifiable.  With the recent
availability of high-quality data for a broad range of performance metrics in
many sports (see, for example, \url{shrpsports.com}), it is now possible to
address questions about measurable aspects of sports competitions that were
inaccessible only a few years ago.  Accompanying this wealth of new data is a
rapidly growing body of literature, both for scientific and lay audiences, on
quantitative modeling and analysis of sports statistics (for general
references, see, e.g., \cite{M97,ABC,KOPR07,AK08,GE09,AM11}).

In this spirit, our investigation is motivated by the following simple
question: can basketball scoring be described by a random walk?  To answer
this question we analyze play-by-play data for four seasons of all National
Basketball Association (NBA) games.  Our analysis indicates that a simple
random-walk model successfully captures many features of the observed scoring
patterns.  We focus on basketball primarily because there are many points
scored per game --- roughly 100 scoring events in a 48-minute game --- and
also many games in a season.  The large number of scoring events allows us
to perform a meaningful statistical analysis.

Our random walk picture addresses the question of whether sports performance
metrics are determined by memory-less stochastic processes or by processes
with long-time correlations (\cite{GVT85,MW91,G96,DC00,EG08}).  To the
untrained eye, streaks or slumps --- namely, sustained periods of superior or
inferior performances --- seem so unusual that they ought to have exceptional
explanations.  This impression is at odds with the data, however.  Impartial
analysis of individual player data in basketball has discredited the notion
of a `hot hand' (\cite{GVT85, AF04}).  Rather, a player's shooting percentage
is independent of past performance, so that apparent streaks or slumps are
simply a consequence of a series of random uncorrelated scoring events.
Similarly, in baseball, teams do not get `hot' or `cold' (\cite{V00,SR09});
instead, the functional forms of winning and losing streak distributions arise
from random statistical fluctuations.

In this work, we focus on the statistical properties of scoring during each
basketball game.  The scoring data are consistent with the scoring rate being
described by a continuous-time Poisson process.  Consequently, apparent
scoring bursts or scoring droughts arise from the Poisson statistics rather
than from a temporally correlated process.  Our main hypothesis is that the
evolution of the score difference between two competing teams can be
accounted by a continuous-time random walk.

This idealized picture of random scoring has to be augmented by two features
--- one that may be ubiquitous and one idiosyncratic to basketball.  The
former is the existence of a weak linear restoring force, in which the
leading team scores at a slightly lower rate (conversely, the losing team
scores at a slightly higher rate).  This restoring force seems to be a
natural human response to an unbalanced game --- a team with a large lead may
be tempted to coast, while a lagging team likely plays with greater urgency.
A similar ``rich get poorer'' and ``poor get richer'' phenomenon was found in
economic competitions where each interaction has low decisiveness
(\cite{DHS98, GS07}).  Such a low payoff typifies basketball, where the
result of any single play is unlikely to determine the outcome of the game.
The second feature, idiosyncratic to basketball, is \emph{anti-persistence},
in which a score by one team is more likely to be followed by a score from
the opponent because of the change in ball possession after each score.  By
incorporating these attributes into a continuous-time random-walk description
of scoring, we build a computational model for basketball games that
reproduces many statistical features of basketball scoring and team win/loss
records.

\section{Scoring Rate}

Basketball is played between two teams with five players each.  Points are
scored by making baskets that are each worth 2 points (typically) or 3
points.  Additional single-point baskets can occur by foul shots that are
awarded after a physical or technical foul.  The number of successive foul
shots is typically 1 or 2, but more can occur.  The duration of a game is
$48$ minutes (2880 seconds).  Games are divided into four 12-minute quarters,
with stoppage of play at the end of each quarter.  The flow of the game is
ostensibly continuous, but play does stop for fouls, time-outs, and out-of-bounds calls.  An
important feature that sets the time scale of scoring is the 24-second clock.
In the NBA, a team must either attempt a shot that hits the rim or score
within 24 seconds of gaining possession of the ball, or else possession is
forfeited to the opposing team.  At the end of the game, the team with the
most points wins.

We analyze play-by-play data from 6087 NBA games for the 2006/07-- 2009/10
seasons, including playoff games (see \url{www.basketballvalue.com}); for
win/loss records we use a larger dataset for 20 NBA seasons
(\url{www.shrpsports.com}).  To simplify our analysis, we consider scoring
only until the end of regulation time.  Thus every game is exactly 48 minutes
long and some games end in ties.  We omit overtime to avoid the complications
of games of different durations and the possibility that scoring patterns
during overtime could be different from those during regulation time.

\begin{figure}[htb]
\begin{center}
\includegraphics[width=0.46\textwidth]{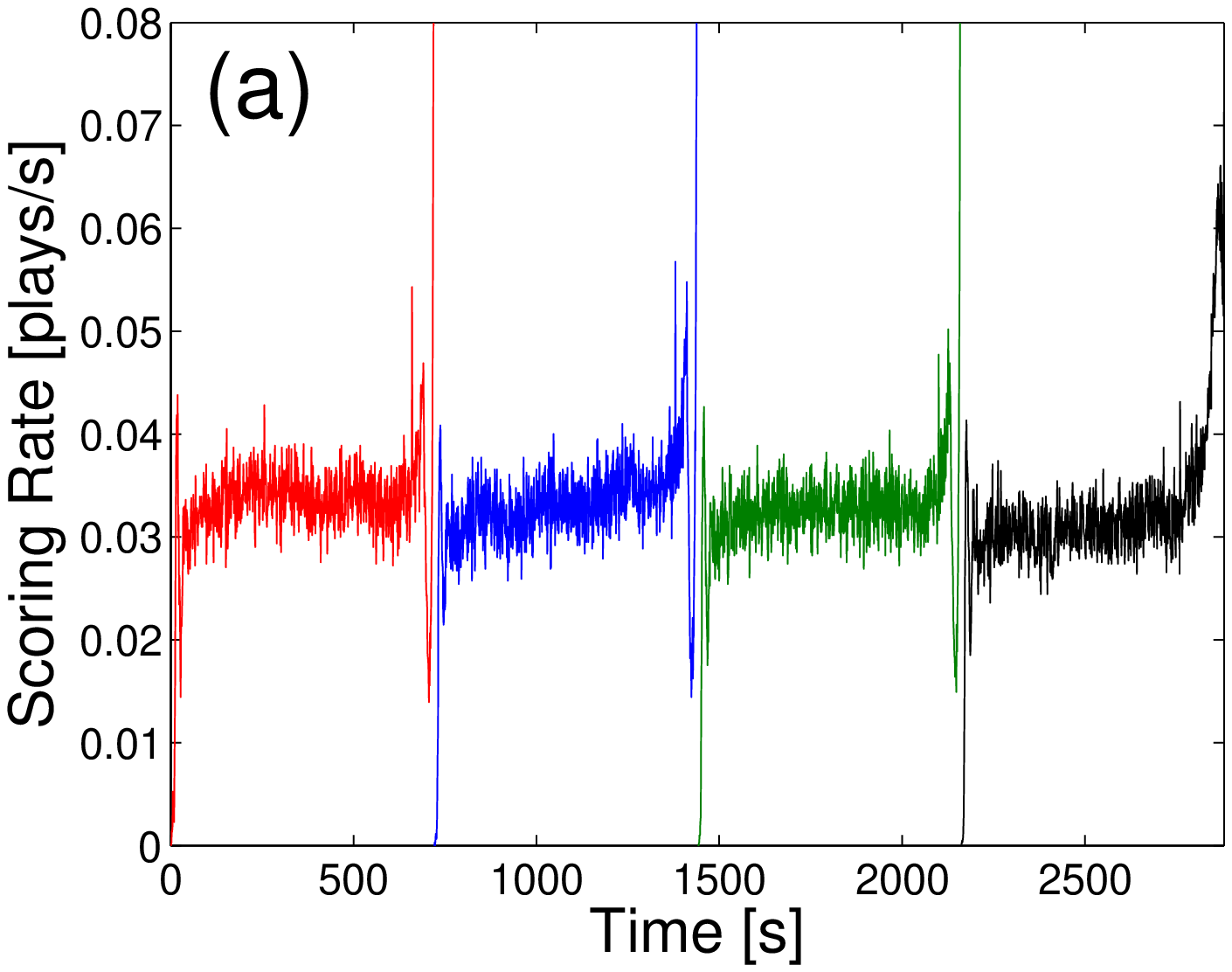} \quad
\includegraphics[width=0.46\textwidth]{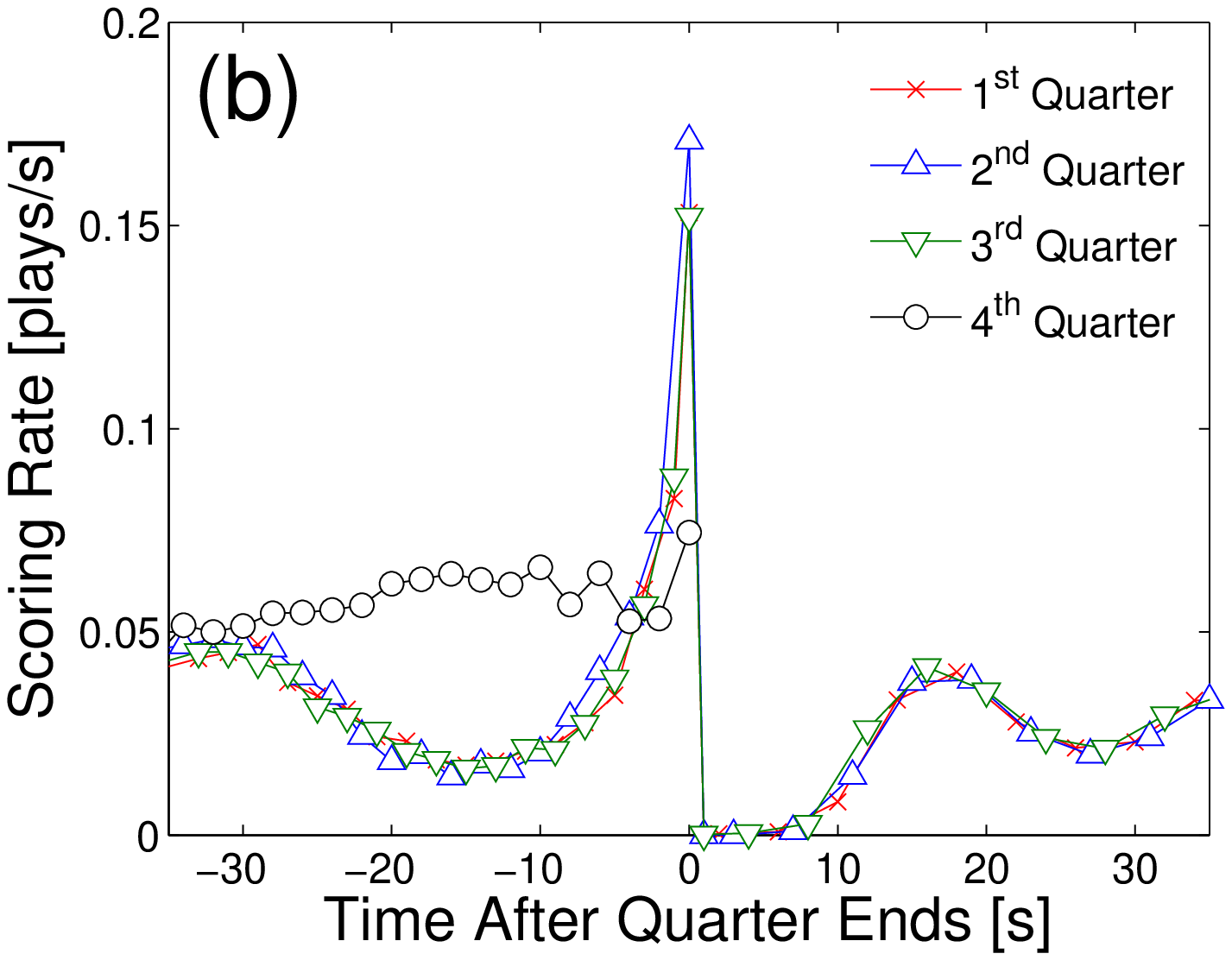}
\caption{(a) Average scoring rate as a function of time over all games in
  our dataset. (b) Rate near the change of each quarter; zero on the
  abscissa corresponds to the start/end of a quarter.}
\label{scoreRate}
\end{center}
\end{figure}

We focus on what we term \emph{scoring plays}, rather than individual
baskets.  A scoring play includes any number of baskets that are made with no
time elapsed between them on the game clock.  For example, a 2-point play
could be a single field goal or two consecutive successful foul shots; a
3-point play could be a normal field goal that is immediately followed by a
successful foul shot, or a single successful shot from outside the 3-point
line.  High-value plays of 5 and 6 points involve multiple technical or
flagrant fouls.  Since they have negligible probability of occurence
(Table~\ref{scoreProb}), we will ignore them in our analysis.  Consistent
with our focus on scoring plays, we define the scoring rate as the number of
scoring plays per second.  This quantity is measured for each
second of the game.  For the 4 seasons of data, the average scoring rate is
roughly constant over the course of a game, with mean value of $0.03291$
plays/sec (Fig.~\ref{scoreRate}).  Averaging each quarter separately gives a scoring rate of 0.03314, 0.03313, 0.03243, and 0.03261 for first through fourth quarters, respectively. The scoring rate corresponds to 94.78 successful plays per
game.  Since there is, on average, 2.0894 points scored per play, each team has
99.018 points in an average game (\cite{graph}).  Parenthetically, the average
scoring rate is constant from season to season, and equals 0.03266, 0.03299,
0.03284, 0.03315 for the 2006--07 to the 2009--10 seasons.

\begin{table}[htb]
\center{\mbox{
\begin{tabular}{|l|l|}
\hline
Points per Basket & Percentage \\
\hline
1 pt. & 33.9\% \\
2 pts. & 54.6\% \\
3 pts. & 11.5\% \\
\hline
\end{tabular}
\quad\quad\quad
\begin{tabular}{|l|l|}
\hline
Points per Play & Percentage \\ 
\hline
1 pt. & 8.70\% \\
2 pts. & 73.86\% \\
3 pts. & 17.28\% \\
4 pts. & 0.14\% \\
5 pts. & 0.023\% \\
6 pts. & 0.0012\% \\
\hline
\end{tabular}
}}
\caption{ Point values of each basket (left) and each play (right) and
  their respective percentages.  }
\label{scoreProb}
\end{table}

Curiously, significant deviations to the constant scoring rate occur near the
start and end of each quarter (Fig.~\ref{scoreRate}(a)).  During roughly the
first 10 seconds of each quarter, scoring is unlikely because of a natural
minimum time to make a basket after the initiation of play.  Near the end of
each of the first three quarters, the scoring rate first decreases and then
sharply increases right at the end of the quarter.  This anomaly arises
because, within the last 24 seconds of the quarter, teams may intentionally
delay their final shot until the last moment, so that the opponent has no
chance for another shot before the quarter ends.  However, there is only an
increase in the scoring rate before the end of the game, possibly because of
the urgent effort of a losing team in attempting to mount a last-minute
comeback via intentional fouls.  While these deviations from a constant
scoring rate are visually prominent, they occur over a small time range near
the end of each quarter.  For the rest of our analysis, we ignore these
end-of-quarter anomalies and assume that scoring in basketball is temporally
homogeneous.

\begin{figure}[ht]
\begin{center}
  \includegraphics[width=0.46\textwidth]{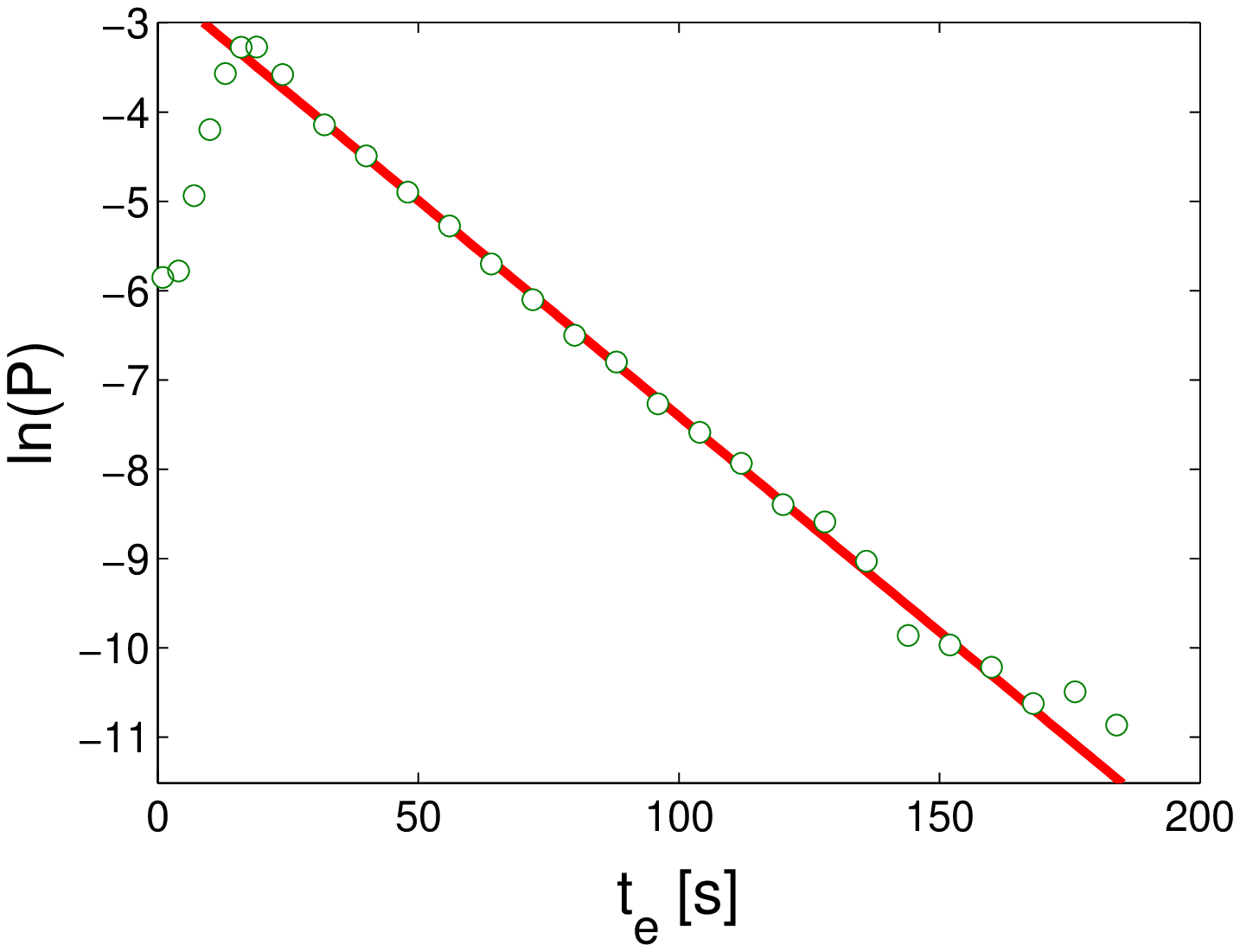}
  \includegraphics[width=0.46\textwidth]{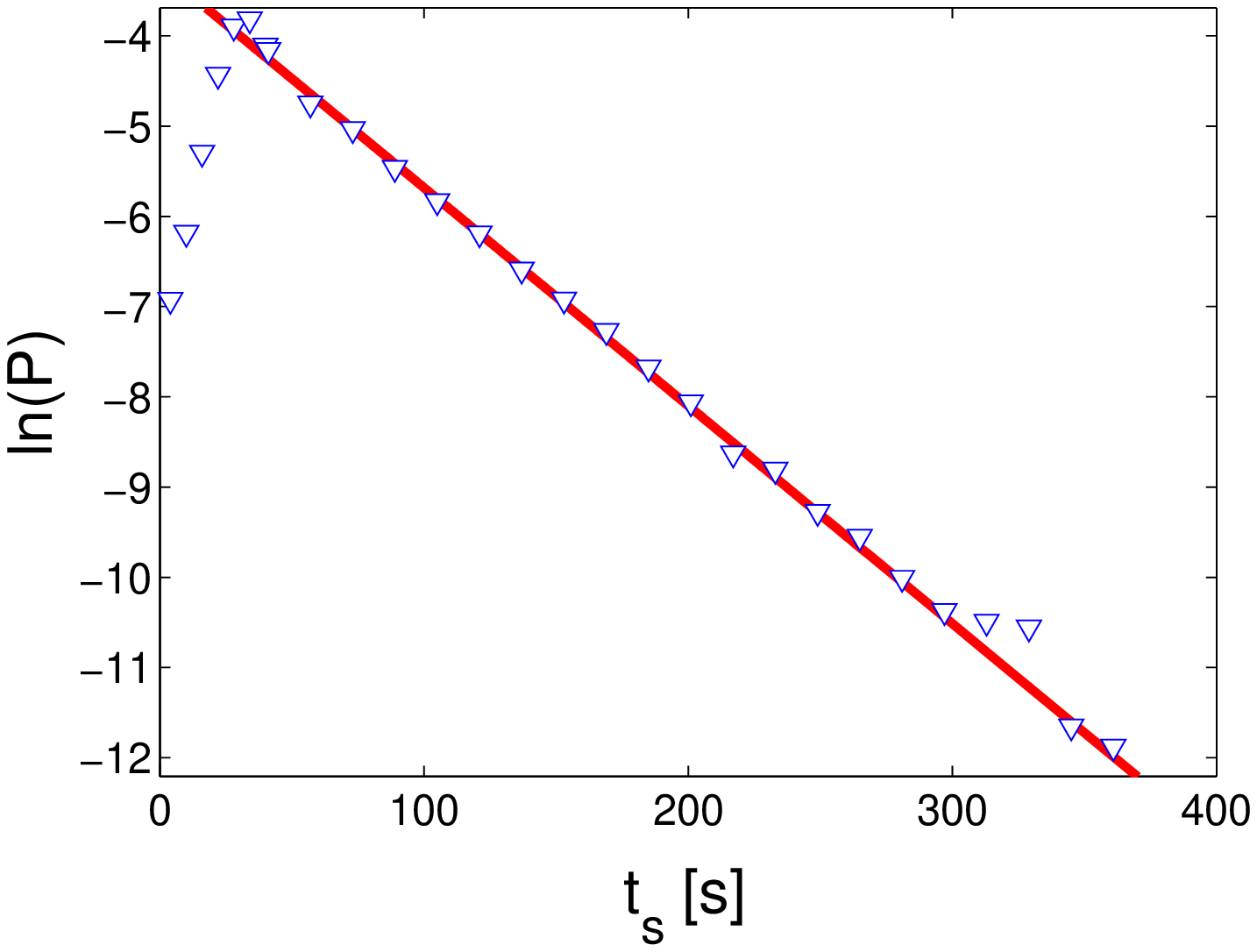}
  \caption{Probability distributions of time intervals between successive
    scores for either team, $P(t_e)$ vs.\ $t_{\rm e}$ (a), and for the same
    team, $P(t_{\rm s})$ vs.\ $t_{\rm s}$ (b).  The line is the least-squares
    linear fit of $\ln(P)$ vs.\ $t$ over the range $t_{\rm e}>30$ sec and
    $t_{\rm s}>60$ sec and corresponds to a decay rate $\lambda_{\rm
      tail}=0.048$ and 0.024, respectively.}
  \label{intervals}
\end{center}
\end{figure}

In addition to temporal homogeneity, the data suggest that scoring frequency
obeys a Poisson-like process, with little memory between successive scores
(see also~\cite{SG11}).  To illustrate this property, we study the
probability $P(t)$ of time intervals between successive scoring plays.  There
are two natural such time intervals: (a) the interval $t_{\rm e}$ between
successive scores of either team, and (b) the interval $t_{\rm s}$ between
successive scores of the same team.  The probability $P(t_{\rm e})$ has a
peak at roughly 16 seconds, which evidently is determined by the 24-second
shot clock.  This probability distribution decays exponentially in time over
nearly the entire range of data (Fig.~\ref{intervals}).  Essentially the same
behavior arises for $P(t_{\rm s})$, except that the time scale is larger by
an obvious factor of 2.  When all the same-team time intervals are divided by
2, the distributions $P(t_{\rm e})$ and $P(t_{\rm s})$ overlap substantially.
The long-time tails of both $P(t_{\rm e})$ and $2P(t_{\rm s}/2)$ are
proportional to the exponential function $\exp(-\lambda_{\rm tail}t)$, with rate
$\lambda_{\rm tail}=0.048$ plays/sec.  This value is larger than the actual
scoring rate of 0.03291 plays/sec because scoring intervals of less than 10
seconds are common for the exponential distribution but are rare in real
basketball games.  Amusingly, the longest time interval in the dataset for
which neither team scored was 402 seconds, while the longest interval for
which a single team did not score was 685 seconds.

\begin{figure}[ht]
\begin{center}
\includegraphics[width=0.6\textwidth]{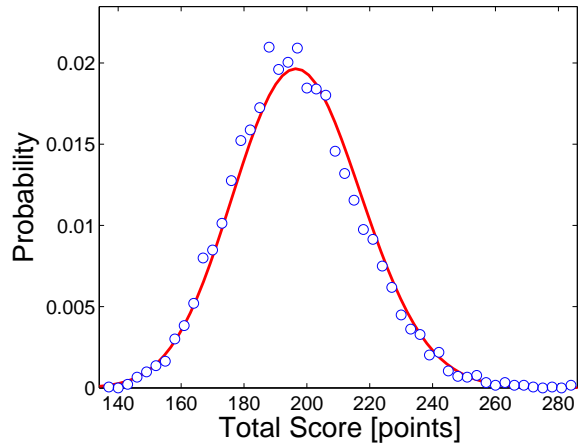} 
\caption{Probability $\mathrm{Prob}({\rm score}=S)$ for a total score $S$ in a single game.  Circles
  are the data, and the solid curve is the Poisson distribution
  \eqref{gamma}. }
  \label{totalScore}
\end{center}
\end{figure}

It is instructive to compare the distribution of total score in a single game
to that of a Poisson process.  Under the assumption that scores occur at the
empirically-observed rate of $\lambda=0.03291$ plays/sec, the probability that
a game has $k$ scoring plays is given by the Poisson distribution,
$\mathrm{Prob}({\rm \#~plays}=k)=\frac{1}{k!}(\lambda T)^ke\,^{-\lambda T}$, where $T=2880$ sec.\ is
the game duration.  Since the average score of each play is $\overline{s}
=2.0894$ points, a game that contains $k$ scoring plays will have a total score
of approximately $S=\overline{s}k$.  By changing variables from $k$ to $S$ in the
above Poisson distribution, the probability that a game has a total score $S$
is
\begin{equation}
\label{gamma}
\mathrm{Prob}({\rm score}=S)= \frac{1}{\overline{s}}\frac{(\lambda T)^{S/\overline{s}}\, 
e^{-\lambda T}}{(S/\overline{s})!}.
\end{equation}
This probability agrees reasonably with game data (Fig.~\ref{totalScore}),
considering that \eqref{gamma} is derived using only the mean scoring rate
and mean points per play.  By including the different point values for each
play, the resulting score distribution would broaden.  Furthermore, if we
impose a cutoff in the probability of short scoring intervals (see
Fig.~\ref{intervals}) the total score distribution of Fig.~\ref{totalScore}
would shift slightly left which would bring the model prediction closer to
the data.

An important aspect of the time intervals between successive scoring events
is that they are weakly correlated.  To illustrate this feature, we take the
time-ordered list of successive scoring intervals $t_1, t_2, t_3, \ldots$, for all games
and compute the n-lag correlation function (\cite{BJ76})
\begin{equation}
\label{corr}
C(n)\equiv \frac{\sum_k (t_k-\overline{t})(t_{k+n}-\overline{t})}{\sum_k (t_k-\overline{t})^2}~.
\end{equation}
Thus $n=1$ gives the correlation between the time intervals between
successive scores, $n=2$ to second-neighbor score intervals, etc.  For both the intervals $t_{\rm e}$
(independent of which team scored) and $t_{\rm s}$ (single team), we find
that $C(n)<0.03$ for $n\geq 1$.  Thus there is little correlation between
scoring events, suggesting that basketball scoring is a nearly memory-less
process.  Accordingly, scoring bursts or scoring droughts are nothing more
than manifestations of the fluctuations inherent in a Poisson process of
random and temporally homogeneous scoring events.

\section{Random-Walk Description of Scoring}

We now turn to the question of \emph{which} team scores in each play to build
a random-walk description of scoring dynamics.  After a given team scores,
possession of the ball reverts to the opponent.  This change of possession
confers a significant disadvantage for a team to score twice in succession.
On average, immediately after a score, the same team scores again with
probability $q=0.348$, while the opponent scores with probability $0.652$.
This tendency for alternating scores is characteristic of an
\emph{anti-persistent\/} random walk (\cite{G07}), in which a step in a given
direction is more likely to be followed by a step in the opposite direction.

\begin{figure}[ht]
\begin{center}
\includegraphics[width=0.6\textwidth]{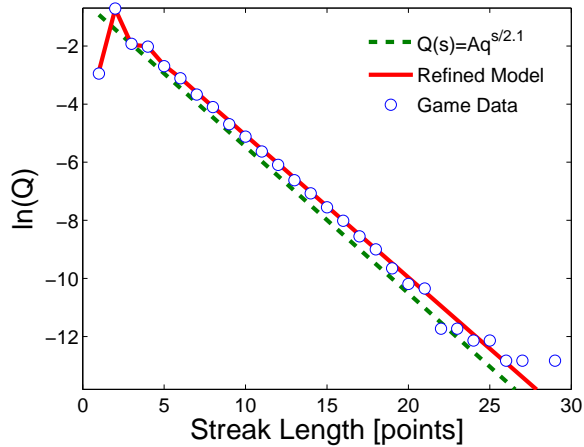} 
\caption{Probability $Q(s)$ for a consecutive point streak of $s$ points
  ($\circ$).  The dashed line corresponds to $Q(s)=Aq^{s/\overline{s}}$,
  with $q=0.348$ and $A$ the normalization constant.  The solid line
  corresponds to a refined model that incorporates the different
  probabilities of 1, 2, 3, and 4-point plays (see Eqs.~\eqref{lowProb} and
  \eqref{recursive}). }
  \label{streaks}
\end{center}
\end{figure}

As we now discuss, this anti-persistence is a determining factor in the
streak-length distribution.  A streak of length $s$ occurs when a team scores
a total of $s$ consecutive points before the opposing team scores.  We define
$Q(s)$ as the probability for a streak to have length $s$.  To estimate this
streak-length probability, note that since $\overline{s}=2.0894$ points are
scored, on average, in a single play, a scoring streak of $s$ points
corresponds to $s/\overline{s}$ consecutive scoring plays.  In terms of an
anti-persistent random walk, the probability $Q(s)$ for a scoring streak of
$s$ points is $Q(s)=Aq^{s/\overline{s}}$ where $A=q^{-1/\overline{s}}-1$ is
the normalization constant.  This simple form reproduces the observed
exponentially decaying probability of scoring streaks reasonably accurately
(Fig.~\ref{streaks}).

However, we can do better by constructing a refined model that incorporates
the different probabilities for 1, 2, 3, and 4 point plays.  Let $w_\alpha$
be the probability that a play is worth $\alpha$ points
(Table~\ref{scoreProb}) and let $v_m$ be the value of the $m^{\rm th}$ play
in a streak.  A scoring sequence $\{v_1,\ldots\,v_n\}$ that results in $s$
points must satisfy the constraint $\sum_{k=1}^n v_k=s$, where $n$ is the
number of plays in the sequence.  The probability for this streak is given by
$\prod_{k=1}^n w_{v_k}$.  Because a streak of length $s$ points involves a
variable number of plays, the total probability for a streak of $s$ points is
\begin{equation}
Q(s)=\sum_{n=1}^\infty \left[ q^{n-1}(1-q) \sum_{\{v_k\}}\left(\prod_{k=1}^n  w_{v_k}\right)\right] \,,
\label{generalPs}
\end{equation}
Here the inner sum is over all allowed sequences $\{v_k\}$ of $n$ consecutive
point-scoring events, and the factor $q^{n-1}(1-q)$ gives the probability for
a streak of exactly $n$ plays.  For example, the probabilities for streaks up
to $s=4$ are:
\begin{align}
\label{lowProb}
\begin{split}
Q(1) &= (1-q)w_1 \\
Q(2) &= (1-q)[w_2 + qw_1^2] \\
Q(3) &= (1-q)[w_3 + 2qw_2w_1 + q^2w_1^3] \\
Q(4) &= (1-q)[w_4 + q(2w_3w_1+w_2^2) + 3q^2w_2w_1^2 + q^3w_1^4].
\end{split}
\end{align}

A direct calculation of these probabilities for general $s$ becomes tedious
for large $s$, but we can calculate them recursively for $s>4$.  To do so,
we decompose a streak of $s$ points as a streak of $s-v_n$ points, followed
by a single play that of $v_n$ points.  The probability of such a play
is $qw_{v_n}$.  Because the last play can be worth 1, 2, 3, or 4 points, the
probability for a streak of length $s$ is given recursively by
\begin{equation}
\label{recursive}
Q(s) = q[w_1Q(s-1) + w_2Q(s-2) + w_3Q(s-3) + w_4Q(s-4)].
\end{equation}
Using Eqs.~\eqref{lowProb} and \eqref{recursive}, we may calculate $Q(s)$
numerically for any $s$.  The resulting probabilities closely match the
empirical data (Fig.~\ref{streaks}), suggesting that streaks arise only from
random statistical fluctuations and not from teams or individuals getting
hot or cold.

\begin{figure}[ht]
\begin{center}
\includegraphics[width=0.6\textwidth]{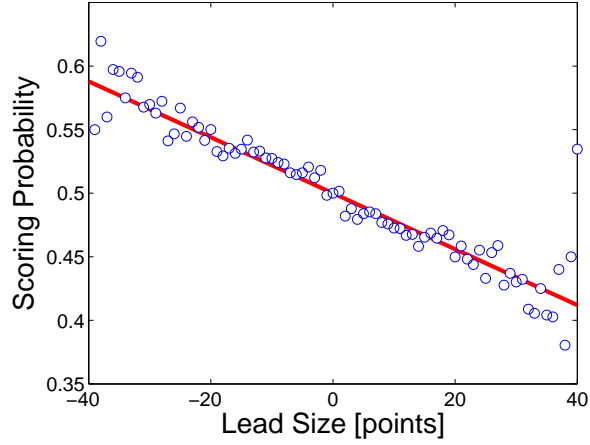} 
\caption{Data for the probability $S(L)$ that a team will score next given a
  lead $L$ ($\circ$).  The line is the least-squares linear fit,
  $S(L)=\frac{1}{2}-0.0022L$.}
  \label{Pvsd}
\end{center}
\end{figure}

Another intriguing feature of basketball games is that the scoring
probability at any point in the game is affected by the current score: the
probability that the winning team scores decreases systematically with its
lead size; conversely, the probability that the losing team scores increases
systematically with its deficit size (Fig.~\ref{Pvsd}).  This effect is
well-fit by a linear dependence of the bias on the lead (or deficit) size.
(Such a linear restoring force on a random walk is known in the physics
literature as the Ornstein-Uhlenbeck model (\cite{UO30}).  For basketball,
the magnitude of the effect is small; assuming a linear dependence, a
least-squares fit to the data gives a decrease in the scoring rate of 0.0022
per point of lead.  Naively, this restoring force originates from the winning
team `coasting' or the losing team increasing its level of effort.

\begin{figure}[ht]
\begin{center}
\includegraphics[width=0.6\textwidth]{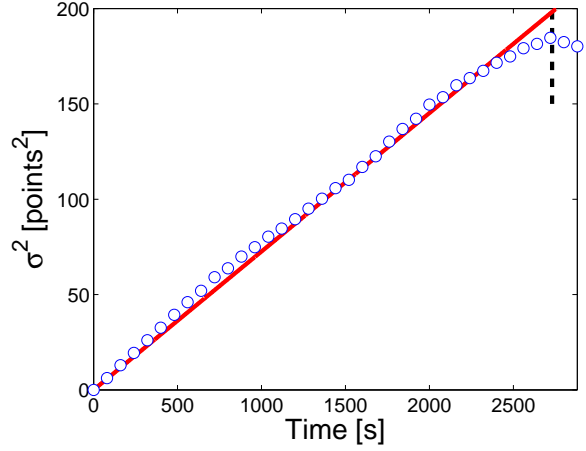} 
\caption{Variance in the score difference, $\sigma^2$, as a function of time.
  The line $\sigma^2=2D_{fit}t$ is the least-squares linear fit, excluding the last 2.5
  minutes of data.  The variance reaches its maximum $2.5$ minutes before the
  end of the game (dashed line). }
  \label{varVSt}
\end{center}
\end{figure}

We now build a random-walk picture for the time evolution of the difference
in the score $\Delta(t)$ between two teams.  Each game starts scoreless and
$\Delta(t)$ subsequently increases or decreases after each scoring play until
the game ends.  The trajectory of $\Delta(t)$ versus $t$ qualitatively
resembles the position of a random walk as a function of time.  Just as for
random walks, the statistically significant quantity is $\sigma^2\equiv {\rm
  var}( \Delta(t))$, the variance in the score difference, averaged over many
games.  For a classic random walk, $\sigma^2=2Dt$, where $D$ is the diffusion
coefficient.  As illustrated in Fig.~\ref{varVSt}, $\sigma^2$ does indeed
grow nearly linearly with time for NBA basketball games, except for the last
$2.5$ minutes of the game; we will discuss this latter anomaly in more detail
below.  A least-squares linear fit to all but the last 2.5 minutes of game
data gives $\sigma^2=2D_{\rm fit}t$, with $D_{\rm fit}=0.0363$
points$^2$/sec.

We may also independently derive an effective diffusion constant from the
time evolution of the score difference from basic parameters of an
anti-persistent random walk.  For such a walk, two successive scores by the
same team correspond to two random-walk steps in the same direction.  As
mentioned above, we found that the probability of this outcome is $q=0.348$.
Conversely, the probability for a score by one team immediately followed with
a score by the opposing team is $1-q$.  Let us define $P(\Delta,t)$ as the
probability that the score difference equals $\Delta$ at time $t$.  Using the
approach of \cite{G07} for an anti-persistent random walk, $P(\Delta,t)$
obeys the recursion
\begin{subequations}
\begin{eqnarray}
  P(\Delta,t+\tau)=q P(\Delta-\ell,t) + q P(\Delta+\ell,t)
  + [(1-q)^2-q^2]P(\Delta, t-\tau),
\label{Difference}
\end{eqnarray}
where $\ell$ is the point value of a single score.  To understand this
equation, we rewrite it as
\begin{eqnarray}
  P(\Delta,t+\tau)= q[P(\Delta-\ell,t)+P(\Delta+\ell,t)-P(\Delta,t-\tau)]
   +(1-q) P(\Delta,t-\tau).
\label{simplify}
\end{eqnarray}
\end{subequations}
The second factor in
\eqref{simplify} corresponds to two scores by alternating teams; thus the
score difference equals $\Delta$ at time $t-\tau$ and again at time $t+\tau$.
This event occurs with probability $1-q$.  The terms in the square bracket
correspond to two successive scores by one team.  Consequently a score
difference of $\Delta\pm2\ell$ at time $t-\tau$ evolves to a score difference
$\Delta$ at time $t+\tau$.  Thus the corresponding walk must be at
$\Delta\pm\ell$ at time $t$ but \emph{not} at $\Delta$ at time $t-\tau$.

Expanding $P(\Delta,t)$ in Eq.~\eqref{Difference} to first order in $t$ and
second order in $\Delta$ yields
\begin{equation}
  \frac{\partial P}{\partial
    t}=\frac{q}{(1-q)}\,\frac{\ell^2}{2\tau}\,\frac{\partial^2 P}{\partial
    \Delta^2}\equiv D_{\rm ap}\,\frac{\partial^2 P}{\partial \Delta^2}~.
\label{TaylorExpand}
\end{equation}
where $D_{\rm ap}$ is the effective diffusion coefficient associated with an
anti-persistent random walk.  Notice that for $q=\frac{1}{2}$ the score
evolution reduces to a simple symmetric random walk, for which the diffusion
coefficient is $D_{\rm ap}=\ell^2/(2\tau)$.  Substituting in the values, from
the game data, $q=0.348$
(probability for the same team to score consecutively), $\ell=2.0894$ (the mean
number of points per scoring event), and $\tau=30.39$ seconds (the average time
between successive scoring events), we obtain
\begin{equation}
  D_{\rm ap} =\frac{q}{1-q}\,\frac{\ell^2}{2\tau}=0.0383\,\,\frac{(\mathrm{points})^2}{\mathrm{sec}}~.
\label{EffectiveD}
\end{equation}
This diffusion coefficient is satisfyingly close to the value $D_{\rm
  fit}=0.0363$ from the empirical time dependence $\sigma^2$, and suggests
that an anti-persistent random-walk accounts for its time dependence.  We
attribute the small discrepancy in the two estimates of the diffusion
coefficient to our neglect of the linear restoring force in the diffusion
equation \eqref{TaylorExpand},

Thus far, we have treated all teams as equivalent.  However, the influence of
team strengths on basketball scoring is not decisive --- weaker teams can
(and do) win against better teams.  The data show that the winning team in
any game has a better season record than the losing opponent with probability
0.6777.  Thus within our random-walk picture, the underlying bias that arises
from the disparity in the strengths of the two competing teams is masked by
random-walk fluctuations.  For a biased random walk with bias velocity $v$
and diffusion coefficient $D$, the competition between the bias and
fluctuations is quantified by the \emph{P\'eclet} number $Pe\equiv v^2t/2D$
(see, e.g., \cite{Pe,R01}), the ratio of the average displacement squared
$(vt)^2$ to the mean-square displacement $2Dt$ caused by random-walk
fluctuations.  For $Pe\ll1$, bias effects due to disparities in team
strengths are negligible, whereas for $Pe\gg1$ the bias is important.  For
basketball, we estimate a typical bias velocity from the observed average
final score difference, $\overline{|\Delta|}\approx 10.7$ points, divided by
the game duration of $t=2880$ seconds to give $v\approx 0.0037$ points/sec.
Using $D\approx 0.0363$ points$^2$/sec, we obtain $Pe\approx 0.55$, which is
small, but not negligible.  Consequently, the bias arising from intrinsic
differences in team strengths is typically not large enough to predict the
outcome of typical NBA basketball games.

\begin{figure}[ht]
\begin{center}
\includegraphics[width=0.8\textwidth]{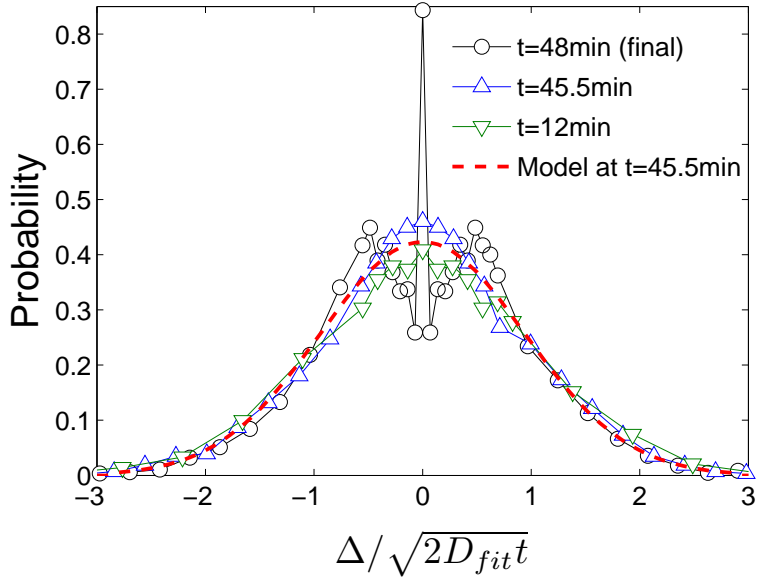} 
\caption{Probability for a given score difference at the end of the first
  quarter, after 45.5 minutes, and at the end of the game.  The abscissa is
  rescaled by linear fit of variance, $\sigma^2\approx2D_{fit}t$ (see Fig.~\ref{varVSt}).  The dashed curve is the
  distribution from simulated games with team strength variance,
  $\sigma^2_X=0.0083$ (see Sec.~4). }
  \label{diffDist}
\end{center}
\end{figure}

Finally, the scoring anomaly associated with the last 2.5 minutes of the game
is striking.  If the score evolves as an anti-persistent random walk, the
distribution of the score difference should be Gaussian whose width grows
with time as $\sqrt{Dt}$.  As shown in Fig.~\ref{diffDist}, the distribution
of score difference has a Gaussian appearance, with a width that grows
slightly more slowly than $\sqrt{Dt}$.  We attribute this small deviation to
the weak restoring force, which gives a diffusion constant that decreases
with time.  However, in the final $2.5$ minutes of the game, the
score-difference distribution develops a spike at $\Delta=0$ and dips for
small $|\Delta|$.  Thus close games tend to end in ties much more often than
expected from the random-walk picture of the score evolution.  This anomaly
may stems from the losing team playing urgently to force a tie, a hypothesis
that accords with the observed increase in scoring rate near the end of the
game (Fig.~\ref{scoreRate}).

\section{Computational Model}

From all of the empirical observations about scoring, we now construct a
computational random-walk model that broadly accounts for point-scoring
statistical phenomena, as well as the win/loss record of all teams at the end
of the season.  In our model, games are viewed as a series of temporally
homogeneous and uncorrelated scoring plays.  The time between plays is drawn
from a Poisson distribution whose mean is the observed value of $30.39$
seconds.  We ignore the short-lived spikes and dips in the scoring rate at
the end of each quarter (Fig.~\ref{scoreRate}) and also the very rare plays
of 5 or 6 points.  Thus plays can be worth 1, 2, 3, or 4 points, with
corresponding probabilities drawn from the observed distribution in
Table~\ref{scoreProb}.  Simulations of scoring events continue until the
final game time of $48$ minutes is reached.

There are three factors that determine \emph{which} team scores.  First, the
better team has a greater intrinsic chance of scoring.  The second factor is
the anti-persistence of successive scoring events that arises from the change
of possession after a score.  The last is the linear restoring force, in
which the scoring probability of a team decreases as its lead increases (and
vice versa for a team in deficit).  We therefore write the probabilities
$P_A$ and $P_B$ that team A or team B scores next, immediately after a
scoring event, as:
\begin{eqnarray}
\label{modelProb}
\begin{split}
  P_A&=I_{A}  - 0.152r -0.0022 \Delta, \\
  P_B&=I_{B} + 0.152r +0.0022 \Delta.
\end{split}
\end{eqnarray}
Here $I_{A}$ and $I_{B}$ are the intrinsic scoring probabilities (which must
satisfy $I_{A}+I_{B}=1$; and the term $\pm 0.152r$ accounts for
the anti-persistence.  Here $r$ is defined as
\begin{equation}
  r= \begin{cases} +1 & \text{team A scored previously},\\
    -1 & \text{team B scored previously},\\
    0 & \text{first play of the game}, \end{cases}
\label{rDef}
\end{equation}
and ensures that the average probability for the same team to score twice in
succession equals the observed value of 0.348.  Finally, the term
$0.0022\Delta$ (with $\Delta$ the score difference) accounts for the
restoring force with the empirically measured restoring coefficient
(Fig.~\ref{Pvsd}).

In our minimalist model, the only distinguishing characteristic of team
$\alpha$ is its intrinsic strength $X_\alpha$.  We estimate team strengths by
fitting simulated team win/loss records to that predicted by the classic
Bradley-Terry competition model (\cite{BT52}), in which the intrinsic scoring
probabilities are given by
\begin{equation}
  I_{A} = \frac{X_A}{X_A + X_B}~, \quad\quad\quad I_{B}=\frac{X_B}{X_A+X_B}~.
\label{pStrengths}
\end{equation}
To simulate a season, we first assign a strength parameter to each team that
is fixed for the season.  We assume that the distribution of strengths is
drawn from a Gaussian distribution with average $\mu_X$ and variance
$\sigma^2_X$ (\cite{JAS93}).  Nearly identical results arise for other team
strength distributions.  Since the intrinsic probabilities, $I_A$ and $I_B$,
depend only on the strength ratio $X_A/X_B$, we may choose $\mu_X=1$ without
loss of generality, so the only free parameter is $\sigma^2_X$.  We determine
$\sigma^2_X$ by simulating many NBA seasons for a league of 30 teams for a
range of $\sigma^2_X$ values and comparing the simulated probability
distributions for various fundamental game observables with corresponding
empirical data.

\begin{figure}[ht]
\begin{center}
\includegraphics[width=0.45\textwidth]{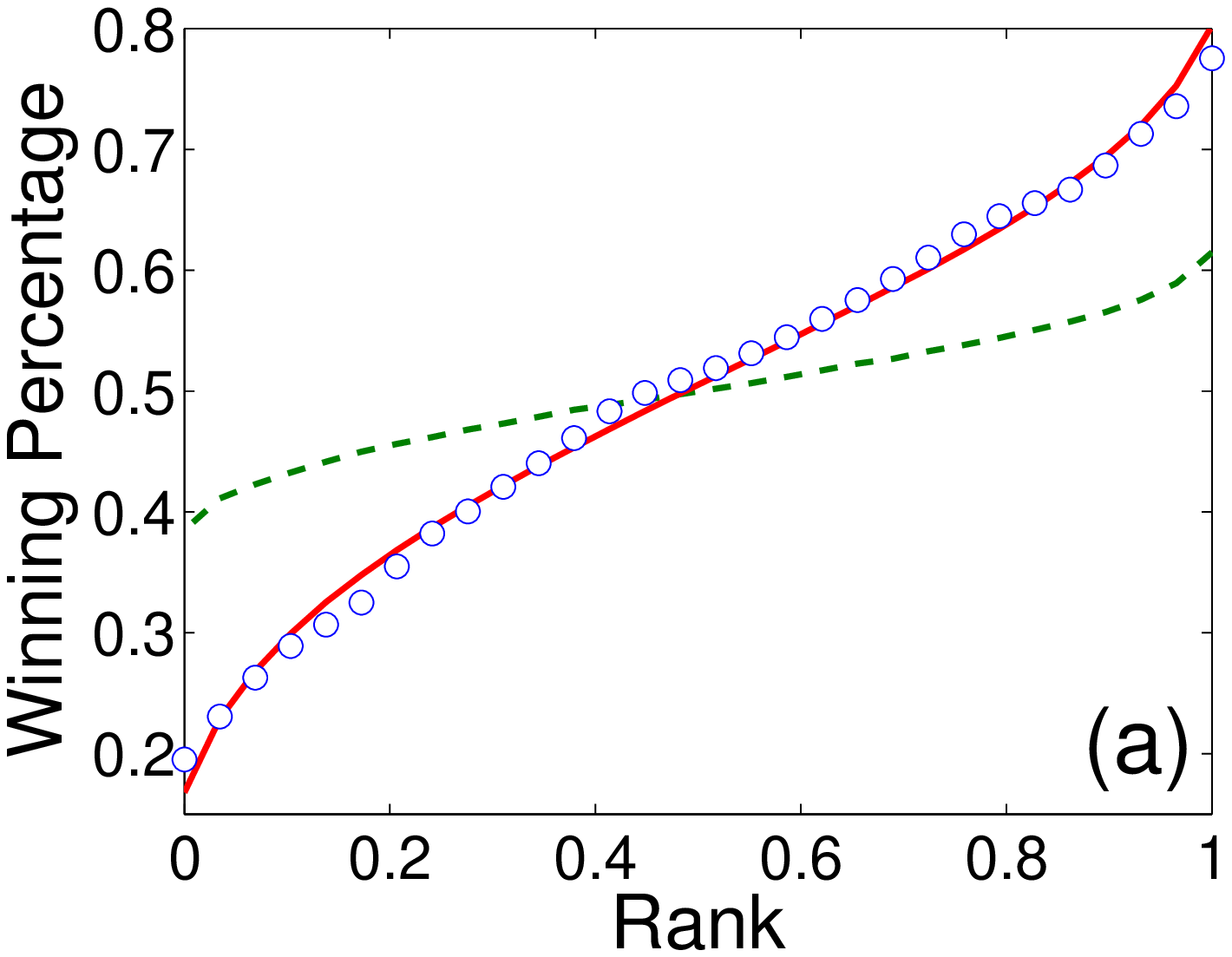} \quad
\includegraphics[width=0.45\textwidth]{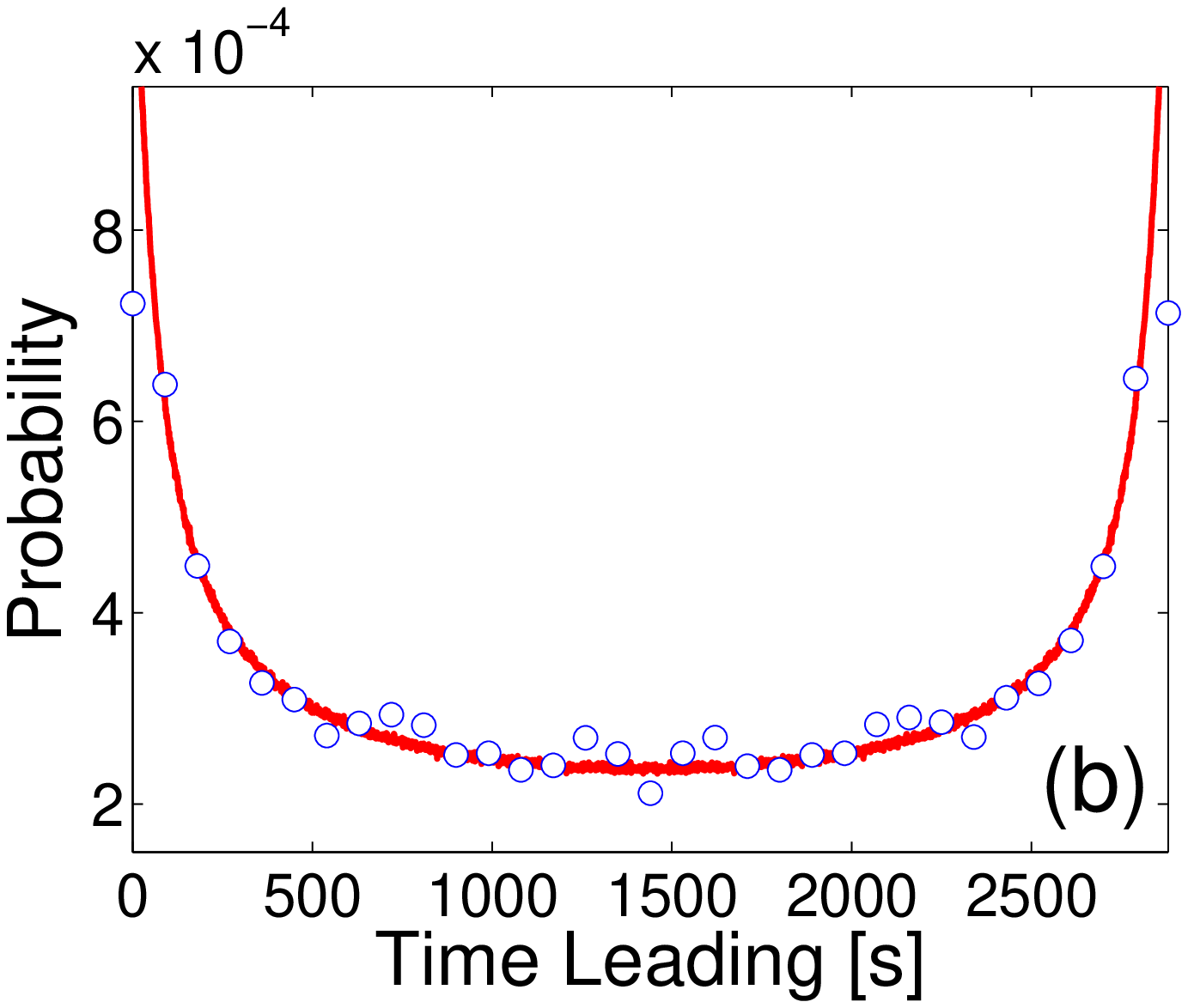} 
\includegraphics[width=0.45\textwidth]{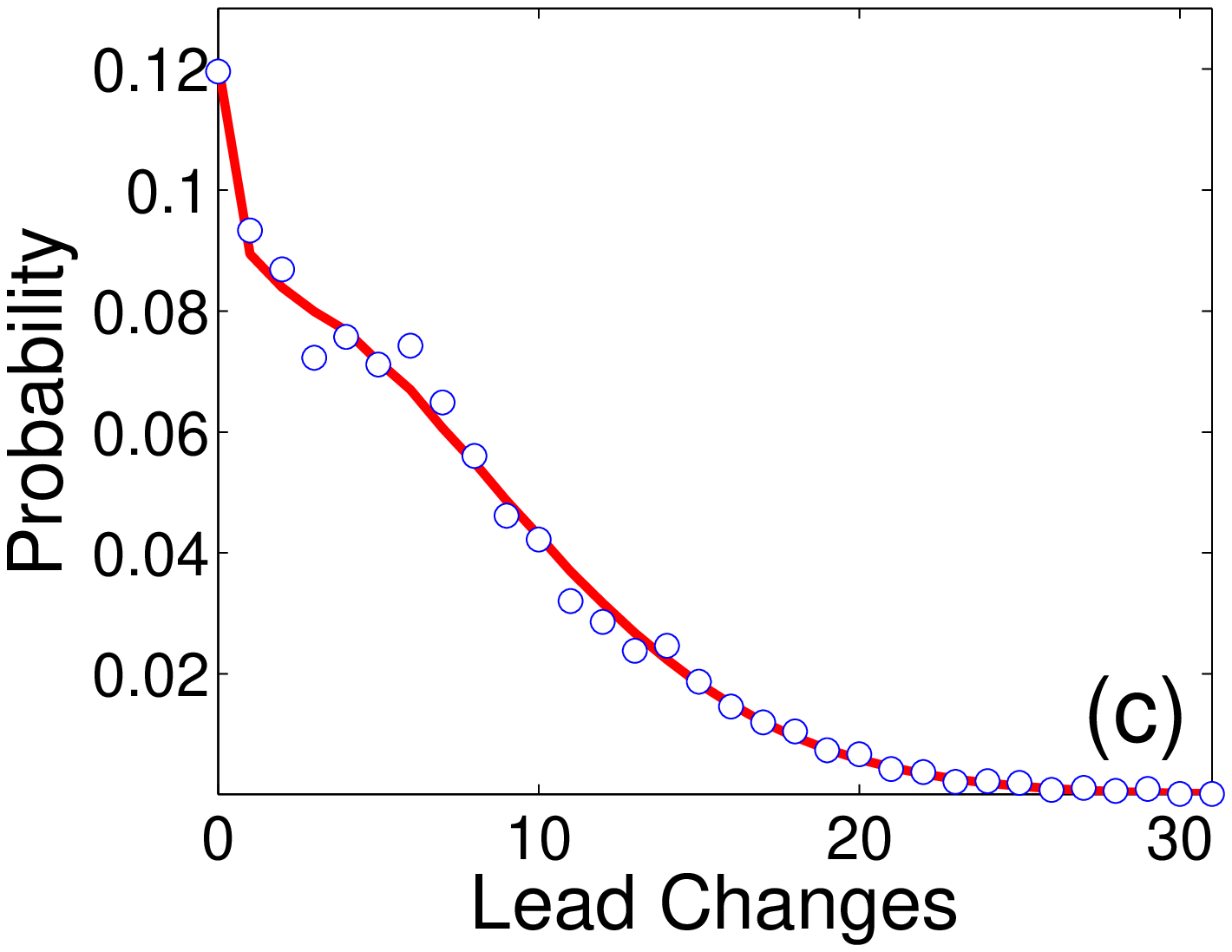}
\caption{(a) Winning percentage as a function of team rank.  The data
  (circles) correspond to the 1991--2010 NBA seasons.  The solid curve is the
  simulated win/loss record when the team strength variance
  $\sigma^2_X=0.0083$.
  The dashed curve is the simulated win/loss record if all teams have equal
  strength, $\sigma^2_X=0$.  (b) Probability that a randomly-selected team
  leads for a given total time. (c) Probability for the number of lead
  changes per game: data ($\circ$) and simulation (curve).  Simulations were
  run for $10^4$ seasons with $\sigma^2_X=0.0083$.}
  \label{ranks}
\end{center}
\end{figure}

Specifically, we examined: (i) The distribution of a given final score
difference (already shown in Fig.~\ref{diffDist}). (ii) The season team
winning percentage as a function of its normalized rank (Fig.~\ref{ranks}
(a)); here, normalized rank is defined so that the team with the best
winning percentage has rank 1, while the team with worst record has rank
0. (iii) The probability for a team to lead for a given fraction of the total
game time (Fig.~\ref{ranks} (b)). (iv) The distribution of the number of
lead changes during a game (Fig.~\ref{ranks} (c)).

Our motivation for focusing on these measures is that they provide useful
statistical characterizations of how basketball games evolve.  The score
difference is the most basic information about the outcome of a basketball
game.  Similarly, the relation between rank and winning percentage provides a
clean overall test of our model.  The probability for a given lead time is
motivated by the well-known, but mysterious arcsine law (\cite{F68}).
According to this law, the trajectory of a one-dimensional random walk is
likely to always be on one side of the origin rather than the walk spending
equal amounts of time to the left and to the right of the origin.  The
ramification of the arcsine law for basketball is that a single team is
likely to lead for the most of the game rather than both teams to equally
sharing the time in the lead.  As a corollary to the arcsine law, there are
typically $\sqrt{N}$ crossings of the origin for a one-dimensional random
walk of $N$ steps, and the distribution in the number of lead changes is
Gaussian.  These origin crossings correspond to lead changes in basketball
games.

For each of the four empirical observables listed above, we compare game data
with the corresponding simulation results for a given value of the team
strength variance $\sigma^2_X$.  We quantify the quality of fit between the
game data and the simulation results by the value $\chi^2$ defined by
\begin{equation}
\chi^2 = \sum_x (F_E(x) - F_S(x))^2~.
\label{chi}
\end{equation}
Here $F_E(x)$ is one of the four above-mentioned empirical observables,
$F_S(x)$ is the corresponding simulated observable, and $x$ is the underlying
variable.  For example, $F_E(x)$ and $F_S(x)$ could be the empirical and
simulated probabilities of the final score difference and $x$ would be the
final score difference.

\begin{figure}[htb]
\begin{center}
\includegraphics[width=0.6\textwidth]{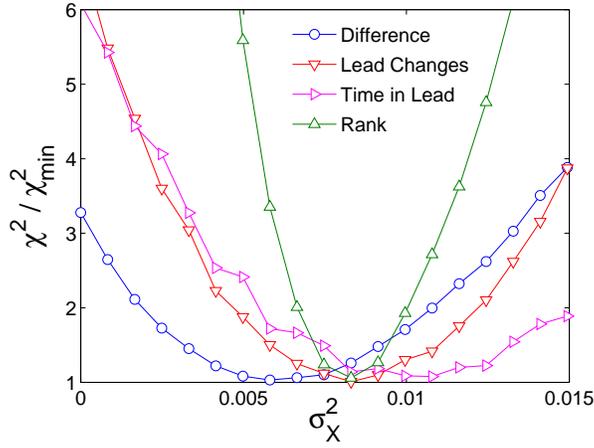} 
\caption{ $\chi^2$ as a function of $\sigma^2_X$ for: the score difference
  distribution at 45.5 minutes ($\circ$), number of lead changes per game
  ($\bigtriangledown$), distribution of time that a team is leading
  ($\triangleright$), and winning percentage as a function of rank
  ($\bigtriangleup$).  Each point is based on simulation of $10^3$ seasons.}
  \label{fitting}
\end{center}
\end{figure}

Figure~\ref{fitting} shows the values of $\chi^2$ as a function of
$\sigma^2_X$ for the four observables.  The best fit between the data and the
simulations all occur when $\sigma^2_X$ is in the range
$[0.00665,\,0.00895]$.  To extract a single optimum value for $\sigma^2_X$,
we combine the four $\chi^2$ measurements into a single function.  Two simple
and natural choices are the additive and multiplicative forms
\begin{eqnarray}
  f_{\rm add}=\sum_{i=1}^{4} \frac{\chi^2_i}{\min(\chi^2_i)}\,, \qquad\qquad
  f_{\rm mult}=\prod_{i=1}^{4} \frac{\chi^2_i}{\min(\chi^2_i)}\,,
\label{combineChi}
\end{eqnarray}
where the sum and product are over the four observables, $\chi^2_i$ is
associated with the $i^{\rm th}$ observable, and $\min(\chi^2_i)$ is its
minimum over all $\sigma^2_X$ values.  The denominator allows one to compare
the quality of fit for disparate functions.  In the absence of any prior
knowledge about which statistical measure about basketball scoring is most
important, we have chosen to weight them equally.  With this choice, both
$f_{\rm add}$ and $f_{\rm mult}$ have minima at $\sigma^2_X=0.0083$.
Moreover, for this value of $\sigma^2_X$, the value of $\chi^2_i$ for each
observable exceeds its minimum value by no more than $1.095$.  These results
suggest that the best fit between our model and empirical data arises when we
choose $\sigma^2_X=0.0083$.  Thus roughly 2/3 of the NBA teams have their
intrinsic strength in the range $1\pm\sqrt{\sigma_x^2}\approx 1\pm 0.09$.



\section{Outlook}
 
From all the play-by-play data of every NBA basketball game over four
seasons, we uncovered several basic features of scoring statistics.  First,
the rate of scoring is nearly constant during a basketball game, with small
correlations between successive scoring events.  Consequently, the
distribution of time intervals between scoring events has an exponential tail
(Fig.~\ref{intervals}).  There is also a scoring anti-persistence, in which a
score by one team, is likely to be followed by a score by the opponent
because of the possession change after each basket.  Finally, there is a
small restoring force that tends to reduce the score difference between
competitors, perhaps because a winning team coasts as its lead grows or a
losing team plays more urgently as it falls behind.  

Based on the empirical data, we argued that basketball scoring data is well
described by a nearly unbiased continuous-time random walk, with the
additional features of anti-persistence and a small restoring force.  Even
though there are differences in the intrinsic strengths of teams, these play
a small role in the random-walk picture of scoring.  Specifically, the
dimensionless measure of the effect of disparities in team strength relative
to stochasticity, the P\'eclet number, is small.  The smallness of the
P\'eclet number means that it is difficult to determine the superior team by
observing a typical game, and essentially impossible by observing a short
game segment.  We simulated our random-walk model of scoring and found that
it satisfyingly reproduces many statistical features about basketball scoring
in NBA games.

This study raises several open issues.  First, is the exponential
distribution of time intervals between scoring events a ubiquitous feature
of sports competitions?  We speculate that perhaps other free-flowing games,
such as lacrosse (\cite{EG08}), soccer (\cite{DC00}), or hockey
(\cite{T07,BWP}), will have the same scoring pattern as basketball when the
time intervals between scores are rescaled by the average scoring rate for
each sport.  It also seems plausible that other tactical metrics, such as the
times intervals between successive crossings of mid-field by the game ball
(or puck) may also be described by Poisson statistics.  If borne out, perhaps
there is a universal rule that governs the scoring time distribution in
sports.

Seen through the lens of coaches, fans, and commentators, basketball is a
complex sport that requires considerable analysis to understand and respond
to its many nuances.  A considerable industry has thus built up to quantify
every aspect of basketball and thereby attempt to improve a team's
competitive standing.  However, this competitive rat race largely eliminates
systematic advantages between teams, so that all that remains, from a
competitive standpoint, are small surges and ebbs in performance that arise
from the underlying stochasticity of the game.  Thus seen through the lens of
the theoretical physicist, basketball is merely a random walk (albeit in
continuous time and with some additional subtleties) and many of the
observable consequences of the game follow from this random-walk description.

\medskip
We thank Guoan Hu for assistance with downloading and processing the data and Ravi Heugel for initial collaborations on this project.  We also thank Aaron Clauset for helpful comments on an earlier version of the manuscript.  This work was supported in part by NSF grant DMR0906504.


%
\bibliographystyle{bepress}

%

%

\end{document}